\documentclass[aps,pra,showpacs, twocolumn, amsfonts, amsmath]{revtex4}

\usepackage{pstricks}

\usepackage{graphicx}

\newcommand{\bra}[1]{\langle #1|}
\newcommand{\ket}[1]{|#1\rangle}

\newcommand{\average}[1]{\langle #1 \rangle}
\newcommand{\dd}{\text{\rm d}}             

\newcommand{\eq}[1]{(\ref{eq:#1})}
\newcommand{\Eq}[1]{Eq.~(\ref{eq:#1})}

\newcommand{\Fig}[1]{Fig.~\ref{fig:#1}}
\newcommand{\fig}[1]{\ref{fig:#1}}

\newcommand{\Sect}[1]{Sect.~\ref{sec:#1}}

\begin{document}


\title{Two-mode Bose gas: Beyond classical squeezing}

\author{C. Bodet$^{1,3}$}
\author{J. Est\`{e}ve$^{2}$}
\author{M. K. Oberthaler$^{2,3}$}
\author{T. Gasenzer$^{1,3}$}
\affiliation{$^{1}$Institut f\"ur Theoretische Physik,
             Ruprecht-Karls-Universit\"at Heidelberg,
             Philosophenweg~16,
             69120~Heidelberg, Germany}
\affiliation{$^{2}$Kirchhoff Institut f\"ur Physik,
             Ruprecht-Karls-Universit\"at Heidelberg,
             Im Neuenheimer Feld 227,
             69120~Heidelberg, Germany}
\affiliation{$^{3}$ExtreMe Matter Institute EMMI,
             GSI Helmholtzzentrum f\"ur Schwerionenforschung GmbH, 
             Planckstra\ss e~1, 
             64291~Darmstadt, Germany} 

\date{\today}

\begin{abstract}
The dynamical evolution of squeezing correlations in an ultracold Bose-Einstein distributed across two modes is investigated theoretically in the framework of the Bose-Hubbard model. 
It is shown that the eigenstates of the Hamiltonian do not exploit the full region allowed by Heisenberg's uncertainty relation for  number and phase fluctuations.
The development of non-classical correlations and relative number squeezing is studied in the transition from the Josephson to the Fock regime.
Comparing the full quantum evolution with classical statistical simulations allows to identify quantum aspects of the squeezing formation.
In the quantum regime, the measurement of squeezing allows to distinguish even and odd total particle numbers.
\end{abstract}

\pacs{03.67.Bg, 
      03.75.-b, 
      03.75.Gg, 
      03.75.Lm, 
      05.20.-y, 	
      05.30.-d  
\hfill HD--THEP--10--04}

\maketitle

\section{Introduction}

Precision measurements at the quantum level are ultimately limited by Heisenberg's uncertainty relation.
Decreasing the fluctuations of an observable of interest below the standard quantum limit given by the central limit theorem necessarily increases the fluctuations in one or more conjugate observables.
Different ways to use such squeezed states, e.g., to measure frequency in Ramsey-type interferometers have been discussed in great detail in the seminal papers \cite{Wineland1992a,Kitagawa1993a,Wineland1994a,Bollinger1996a}.
Heisenberg-limited Mach-Zehnder interferometry using number squeezed photon states were studied in detail in \cite{Holland1993a,Kim1998a,Huelga1997a,Dunningham2002b}.
The standard quantum limit is reached by today's best sensors of various quantities such as time \cite{Santarelli1999a} and position \cite{Arcizet2006a,Goda2008a}.

\begin{figure}[htb]
\begin{center}
\includegraphics[width=0.45 \textwidth ]{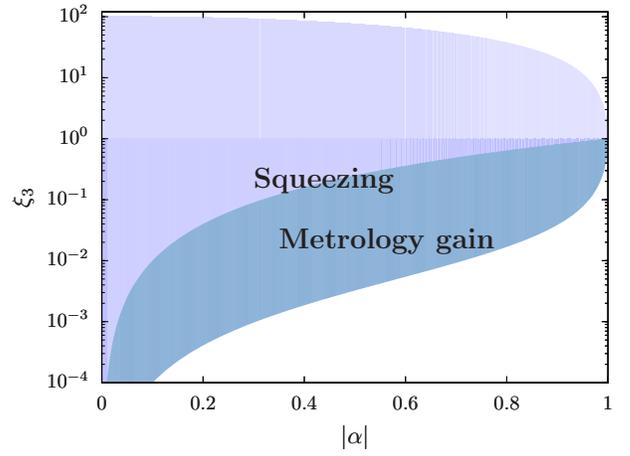}
\caption{ \label{fig:PhaseDiagIntro}
(Color online)
Phase diagram for $N=100$ particles in a double-well potential, with on the average equal populations in the two wells, $\average{\hat n}=\average{\hat n_{1}-\hat n_{2}}/2=0$, and the absolute phase chosen such that $\average{\hat S_{2}}=i\average{{\hat a}_{2}^\dagger{\hat a}_{1}-{\hat a}_{1}^\dagger{\hat a}_{2}}=0$, see \Eq{Jdefine}. 
For any given coherence parameter $\alpha=2\average{\hat S_{1}}/N$ the variance $(\Delta n)^2=N\xi_{3}/4$ of the number difference between the wells is bounded below and above by Heisenberg's uncertainty relation and the constraint that the total particle number $N$ is fixed.
The allowed area is defined by the inequality \eq{xi31bound} shaded in colors in the above diagram.
In the medium (violet) and dark (blue) shaded regimes, the squeezing parameter is below the standard quantum limit, $\xi_{3}\le1$, see \Eq{xi3}.
In the dark (blue) regime, the squeezing can be used to gain precision in metrology, see \Eq{entanglement}.
}
\end{center}
\end{figure}
Two-mode Bose-Einstein condensates of non interacting particles constitute coherent semiclassical macroscopic ensembles of particles.
In the limit of zero temperature the probability for each mode to contain a certain number of atoms is approximately Poissonian.
The variance $(\Delta n)^{2}$ of the particle number difference is proportional to the total particle number $N$, corresponding to the standard quantum limit $(\Delta n)/N\sim 1/\sqrt{N}$.
Conjugate to particle number difference is the relative phase $\phi$ between the modes which can be measured through interference effects
\footnote{We note that both, phase and particle number in a closed non-relativistic system of massive particles can only be measured relative to that of a different system, or they can be measured locally comparing the system at different space-time points. 
In a closed system a finite mass implies a fixed total particle number and therefore an undefined total phase.}.
The Heisenberg uncertainty relation $\Delta n\Delta\phi\sim 1$ implies the relative standard deviation $(\Delta\phi)/\phi\sim 1/\sqrt{N}$. 

The presence of interactions between the particles strongly modifies the situation.
Fluctuations of the particle number and squeezing in trapped atomic gases have been the subject of numerous recent experimental and theoretical studies 
\cite{Orzel2001a,Greiner2002b,Gerbier2006a,Sebby-Strabley2007a,Jo2007a,Li2007a,Esteve2008a}.
Here we are interested in squeezed states of Bose-Einstein condensed ensembles of trapped atoms.
We specifically consider squeezing in the particle number difference between the two minima of a double-well trap in one spatial dimension.
This reduction of number fluctuations below the standard quantum limit occurs at the expense of increased fluctuations in the relative phase between the wells.
The barrier between the wells of the system is taken much smaller than the outer walls of the trap such that the main source of number fluctuations in each well is given by tunneling processes between the wells.
Such two-mode quantum systems can be described, by use of Schwinger's representation, in terms of angular momentum states with the maximum length of the spin vector related to the total particle number.
Consequently, the non-classical states we consider exhibit a variant of spin squeezing \cite{Wineland1992a,Kitagawa1993a}. 

Our studies have been initiated by a recent experiment \cite{Esteve2008a} in which a setup of the kind sketched above has been used, with a one-dimensional double-well trap created by a superposition of standing light waves.
In this experiment, both particle number and phase difference have been measured independently with high resolution to establish squeezing correlations empirically.
While our discussion of the formation of squeezed states here focuses on particle number variations in space, it can be straightforwardly extended to apply to alternative schemes, including, e.g., squeezing in the relative occupation number of internal hyperfine states of the atoms \cite{Gross2009a}.  

A prominent motivation of the work presented here is to study the role of quantum statistical fluctuations in the preparation of the spin squeezed states.
By comparison of exact quantum with semi-classical Monte Carlo simulations it is shown here that the spin squeezing produced in a setup as that in the experiment \cite{Esteve2008a} is equivalent to reduced classical fluctuation of the occupation number difference between the two wells.
It is, furthermore shown that the production of squeezed states in the experiment follows a quasi static path in state space. 

Spin squeezing is closely related to quantum entanglement \cite{Braunstein1999a,Simon2000a,Duan2000c,Sorensen2001a,Sorensen2001b,Wang2003a,Korbicz2005a}.
Schemes have been proposed to use spin-squeezed states for quantum teleportation of continuous-variable, e.g. coherent states \cite{Bennett1993a,Vaidman1994a,Braunstein1998a}. 
However, a full reconstruction of the teleported state is only possible in the limit of perfect squeezing \cite{Braunstein1998c}, close to zero temperature where quantum fluctuations become relevant, beyond the regime of classical squeezing.
We demonstrate to which extent the evolution of the system enters the regime of non-classical states.

We furthermore show that, for a finite total number of atoms the Heisenberg limit can not be reached, in the generic case, by adiabatically changing the system's parameters, starting from an incoherent mixture of energy eigenstates.
In particular, in  the ground state of the system, though being rather close to the Heisenberg limit,  the variances of the conjugate variables number difference and relative phase between the wells are in general not minimized at the Heisenberg limit.
The difference between ground state and Heisenberg limited variances becomes negligible only in the limit of large total particle number.
Any excited energy eigenstate is found to be even further from this limit.
The production of states at the Heisenberg limit therefore requires nonequilibrium dynamical evolution.

Our paper is organized as follows:
In \Sect{StatSys} we discuss fundamental squeezing limits in the Bloch-sphere picture.
The dynamics of the production of squeezing by raising the barrier in the double-well potential is described in \Sect{Dynamics}.
A focus is set on comparing the full quantum with the semiclassical evolutions.
Our conclusions are drawn in \Sect{Concl}.

\section{Number-squeezing in a two-mode Bose gas}
\label{sec:StatSys}
We consider a Bose condensate trapped in a double-well potential with two energetically
degenerate minima separated by a barrier of variable height, allowing for an
adjustable tunneling rate between the wells.
In the experiment \cite{Esteve2008a} such a potential was formed
optically by counterpropagating laser waves.
We consider parameter regimes in which the system can be described by the two-site Bose-Hubbard Hamiltonian
\begin{equation}
\label{eq:BHeqn}
  \hat H 
  = - J (\hat a_1^\dagger \hat a_{2}+\hat a_{2}^\dagger \hat a_1) 
     + \frac{U}{2} \sum_{i=1}^2 \hat a_i^\dagger \hat a_i^\dagger \hat a_i \hat a_i.
\end{equation}
where the operators $\hat a_i$ and $\hat a_i^\dagger$ obey the standard bosonic commutation relations
\begin{equation}
  \label{eq:commutation}
  [\hat a_i, \hat a_j]=0,\quad
  [\hat a_i^\dagger, \hat a_j^\dagger]=0,\quad
  [\hat a_i, \hat a_j^\dagger]=\delta_{ij}.
\end{equation}
and where $J$ is the tunneling and $U$ is the onsite interaction parameter.
The energy spectrum and eigenstates of this Hamiltonian, as far as relevant for the discussion in this article, are summarized in Appendix \ref{app:StatSys}.
The Hamiltonian can be written in terms of Schwinger angular momentum operators,
\begin{equation}
\label{eq:HamilAngular}
  \hat H = - J (\hat S_{+}+\hat S_{-}) + U (\hat S_{3}^2 + \hat {\mathbf S}^{2}-N),
\end{equation}
where $\hat S_\pm=\hat S_{1}\pm i\hat S_{2}$, and $\hat {\mathbf S}^{2}=\sum_{i=1}^{3}\hat S_i^{2}$, with $\hat S_{1}=(\hat a_2^{\dagger}\hat a_1+\hat a_1^{\dagger}\hat a_2)/2$, $\hat S_{2}=i(\hat a_2^{\dagger}\hat a_1-\hat a_1^{\dagger}\hat a_2)/2$, and $\hat S_{3}=(\hat a_1^{\dagger}\hat a_1-\hat a_2^{\dagger}\hat a_2)/2$.
The corresponding angular-momentum-type states of the system can be represented by quantum phase-space (Wigner) distributions on the Bloch sphere \cite{Dowling1994a,Barnett1997a}.
For details about this representation we defer to Appendices \ref{app:StatSys} and \ref{app:Wigner}. 

The difference in occupation number between the modes is measured by $\hat S_{3}$, while the relative phase coherence is measured by the orthogonal components $\hat S_{1,2}$. 
These angular-momentum-type observables are constraint by a Heisenberg uncertainty relation, see \Eq{UncertaintyS}, which allows us to derive fundamental limits for squeezing.

To quantify angular momentum squeezing one introduces a squeezing parameter adapted to the problem under consideration.
A definition suitable for number squeezing between bosonic modes relates the variance of one Cartesian component to the total spin $S=N/2$, i.e., to the total number of mode excitations or particles, 
\begin{align}
  \xi_{i}=(\Delta S_{i})^2/(S/2).
\end{align}
For $i=3$, this is equivalent to the ratio of the variance of the number difference $n_{1}-n_{2}=2n$ to the total number,
\begin{align}
  \xi_{3}=4(\Delta n)^2/N.
  \label{eq:xi3}
\end{align}
In the following section, we show the evolution of this squeezing parameter under the slow ramp-up of the potential barrier.
It essentially reflects the suppression of the relative number fluctuations below the classical limit $(\Delta n)^2= N/4$ as given by the central-limit theorem.

In the following we will consider only situations where the total number $N$ of atoms in the wells and therefore the total spin $S$ are fixed.
Before proceeding with this we briefly remark that in the general case of varying $S$ the squeezing parameters $\xi_{i}$ measure the deviation from an angular momentum coherent state which is represented, in Bloch space, as a spherical Gaussian uncertainty distribution around $\langle \mathbf{S}\rangle$ with radial width $\sigma=\sqrt{N}/2$.
Such a state can be written as a product of coherent states in the two modes 1 and 2, $|\alpha\rangle|\alpha\rangle$, $|\alpha|=\sqrt{N/2}$, and has $\xi_{1}=\xi_{2}=\xi_{3}=1$.

%
\begin{figure}[tb]
\begin{center}
\includegraphics[width=0.48 \textwidth ]{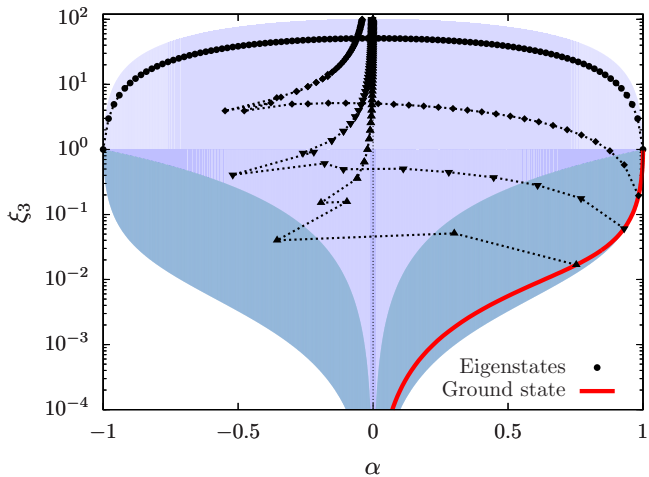}
\caption{ \label{fig:PhaseDiagEEigenstates}
(Color online)
Positions of the energy eigenstates of the Hamiltonian \eq{BHeqn} for a system with $N=100$ particles in the phase diagram introduced in \Fig{PhaseDiagIntro}.
Each black dot corresponds to one such state.
The dotted lines are drawn to guide the eye between successive eigenstates of a Hamiltonian with fixed ratio $U/J$. 
From top to bottom, the sets of states are obtained for $U/J=0, 1, 10$, and $100$, respectively.
The (red) solid line connects all ground states for different $U/J$.
We emphasise that the energy eigenstates do not extend over the full (shaded) region allowed by Heisenberg's uncertainty.
}
\end{center}
\end{figure}
%
In analogy to spin-squeezed states, two-mode states can exhibit squeezing in particular directions at the expense of increased fluctuations in directions perpendicular to this \cite{Wineland1994a}. 
For example, an eigenstate of $\hat S_3$ can be illustrated by a circle on the Bloch sphere parallel to the $1$-$2$-plane corresponding to the variances
\begin{align}
\begin{split}
  (\Delta S_1)^2=(\Delta S_2)^2 
  &=[S(S+1)-S_{3}^{2}]/2\\
  (\Delta S_3)^2 
  &=0,
\end{split}
\end{align}
respectively.
The operators $\hat S_1$, $\hat S_2$ measure the phase between the Fock modes $1$ and $2$ which is accessible to interference measurements of the particle occupation numbers, see Ref.~\cite{Esteve2008a}. 
Hence, the commutators \eq{angular} characterise the uncertainty relation \eq{UncertaintyS} between (relative) number and cosine of the (relative) phase.
The absolute phase is undefined as we assume a fixed total particle number. 
Physically this phase is not measurable without comparing and therefore coupling the system to another system.
Hence, the two-mode states considered in the following, with relative particle number centered around $n=0$, correspond to a distribution on the surface of the Bloch sphere, centered around the equator, i.e., around a polar angle $\theta=\pi/2$ or $\average{\hat S_3}=0$. 
A more convenient parameter to reflect the squeezing below the standard quantum limit, in accordance with the uncertainty relation \eq{UncertaintyS} is given by
\begin{align}
  \xi_{kl} = (\Delta S_{k})^2/|\average{\hat S_{l}}/2|.
\end{align}
In the following, we will consider the special case $\langle\hat  S_{2}\rangle=\langle\hat  S_{3}\rangle=0$ such that the only nontrivial combinations are $\xi_{21}$ and $\xi_{31}$.
The variances $(\Delta S_{k})^2$ are subject to the uncertainty relation
\begin{align}
\label{eq:UncertaintyS}
  (\Delta S_{k})^2(\Delta S_{l})^2 \ge \frac{1}{4} |\epsilon_{klm}\langle \hat S_{m}\rangle|^{2}.
\end{align}
such that the parameters $\xi_{kl}$ obey the inequality
\begin{align}
\label{eq:Uncertaintyxikl}
  \xi_{21}\xi_{31} \ge 1.
\end{align}
This defines the Heisenberg limit for the fluctuations.
One may at first glance expect that the limit $\xi_{31}\to 0$ is allowed at the expense of $\xi_{21}\to\infty$ and vice versa.
However, as long as $|\average{\hat S_{1}}|>0$ perfect squeezing in either the relative number or the phase is not possible as we have assumed fixed $S$ and thus a fixed Bloch-sphere radius.

In order to determine the actual squeezing limit for a fixed total number it is, moreover, not sufficient to take the general upper limit $(\Delta S_{2})^2\le S^{2}=N^2/4$ and infer, from the uncertainty relation \eq{UncertaintyS} for the variances of the angular momenta, the lower limit $(\Delta S_{3})^2\ge\average{\hat S_{1}}^2/N^2$, which vanishes at $\average{\hat S_{1}}=0$ and becomes as large as $1/4$ at $\average{\hat S_{1}}=N/2$.

Since the variance of $S_{2}$ depends nontrivially on the mean value of $S_{1}$ it is rather necessary to take into account the constraint given by the fixed total particle number $N=2S$.
Using
\begin{align}
  \average{{\hat {\mathbf S}}^{2}}=S(S+1)
\end{align}
one finds, from \eq{Uncertaintyxikl}, the inequality
\begin{align}
\label{eq:xi31bound}
  &\xi_{31}^{2}-2\xi_{31}\gamma+1 \le 0,
  \\
  \label{eq:gamma}
  \mbox{with}\quad
  &\gamma=\frac{S(S+1)-\average{\hat S_{1}^2}}{|\average{\hat S_{1}}|}.
\end{align}
This contains, besides the average, also the fluctuations of the ``phase operator'' $\hat S_{1}$.
In general the above inequality returns a lower bound $\xi_{31,\mathrm{min}}\le\xi_{31}$ and an upper bound $\xi_{31,\mathrm{max}}\ge\xi_{31}$ for $\xi_{31}$.

We show the resulting bounds on $\xi_{3}=4(\Delta n)^2/N=4(\Delta S_{3})^2/N=2\xi_{31}|\average{\hat S_{1}}|/N$ as functions of the coherence $\alpha=2\average{\hat S_{1}}/N$ in \Fig{PhaseDiagIntro}.
The shaded areas are allowed by the inequality \eq{xi31bound}.
In the medium (violet) and dark (blue) shaded areas the system is below the standard quantum limit, $\xi_{3}\le1$, see \Eq{xi3}.

For the special case that $\average{\hat S_{1}}=S=N/2$, i.e., $\alpha=1$, one has $\average{\hat S_{1}^2}=S^2$ such that the inequality \eq{xi31bound} becomes an equation fixing the variance to the unique value $(\Delta S_{3})^2= N/4$ which is both a lower and an upper bound.
Note that this bound is enhanced by a factor $N$ as compared to the naive limit derived from the uncertainty relations above.
For $|\average{\hat S_{1}}|<N/2$ one finds, since $\average{\hat S_{1}^2}\le S(S+1)$, that $\gamma\ge1$. 
Moreover, $\average{\hat S_{1}^2}\ge\average{\hat S_{1}}^2$, and the minimum $\average{\hat S_{1}^2}=\average{\hat S_{1}}^2$ is realized, for any $\average{\hat S_{1}}$, by the eigenstates of $\hat S_{1}$.
Hence, one can replace $\average{\hat S_{1}^2}$ by $\average{\hat S_{1}}^2$ in \eq{gamma}, such that the inequality \eq{xi31bound} defines the lower and upper bounds to $\xi_{31}$ for any given value of the average $\average{\hat S_{1}}$. 
For $|\average{\hat S_{1}}|\to0$, one finds a lower bound, to quadratic approximation in $\average{\hat S_{1}}$, of \footnote{%
This inequality was also found in Ref.~\cite{Sorensen2001b}, where Eq.~(3), taking into account that the square-bracketed term under the root needs to be squared, gives our bound in the limit of small $\average{J_{z}}=\average{\hat S_{1}}$, see arXiv: quant-ph/0011035v2.}

\begin{align}
\label{eq:Delta3boundlowS1}
  &(\Delta S_{3})^2\ge\frac{\average{\hat S_{1}}^2}{4S(S+1)},
\end{align}
This limit is by a factor of $S/(S+1)$ lower than the naive bound mentioned above, a difference which only disappears in the limit of large particle numbers.
Also the maximally allowed $(\Delta S_{3})^2$ is given by $S(S+1)$ instead of $S^2$.

The angular momentum representation \eq{HamilAngular} of the Hamiltonian shows that, for a given set of parameters $U$ and $J$, each of the energy eigenstates corresponds to a point in the $|\average{\hat S_{1}}|$-$(\Delta S_{3})^2$ plane.
In \Fig{PhaseDiagEEigenstates} we show these points for $J=1$~Hz and four different values of $U$ (black dots). 
Those values are, from top to bottom, $U/J=0, 1, 10$, and $100$. 
The dotted lines serve to guide the eye between the values for subsequent states of the same Hamiltonian.
This shows that states with lower energies have larger $|\average{\hat S_{1}}|$ and smaller $(\Delta S_{3})^2$ than states with higher energy.
We find that the ground-state values are in accordance with the limit set by the Heisenberg uncertainty relation \eq{xi31bound} but correspond with this limit only for coherences $\alpha=0$ and $\alpha=1$.
Similarly, the eigenstates with the highest energies, do not give the highest possible number fluctuations $(\Delta S_{3})^2$.
The only exception are the cases $\average{\hat S_{1}}=\pm S$, i.e., $U=0$, where the lower and upper Heisenberg limits meet at $\xi_{3}=1$, and the case $\average{\hat S_{1}}=0$.
Hence, the diagram in \Fig{PhaseDiagEEigenstates} shows that adiabatic changes of the parameters cannot drive the system, starting in some classical (diagonal) mixture of energy eigenstates, into the maximally squeezed state, for any value of the coherence $\alpha$ other than $0$ or $\pm1$.
We emphasise that maximum squeezing allowed by Heisenberg's uncertainty relation is only possible in a non-equilibrium procedure.

Spin-squeezed states with large total angular momentum quantum number $S$ have been suggested as means for increasing the precision of interferometric and metrology measurements beyond the standard quantum limit \cite{Wineland1992a,Kitagawa1993a,Wineland1994a,Bollinger1996a}.
One thereby uses, e.g., the squeezing of the uncertainty ellipsoid around the mean spin vector $\average{\hat{\mathbf{S}}}$. If the ellipsoid is squeezed perpendicular to the spin direction along a direction $\mathbf{\sigma}$ this increases the measurement sensitivity of an angle $\theta$ of rotation of $\average{\hat{\mathbf{S}}}$ about an axis ${\mathbf \rho}$ perpendicular to the spin and the squeezing direction, $\average{\hat{\mathbf S}}\cdot{\mathbf \rho}={\mathbf \sigma}\cdot{\mathbf \rho}=0$.
The resolution of the angle $\theta$ is proportional to the variance $(\Delta{\mathbf{S}}_{\mathbf{\sigma}})^2$ of the spin vector along the squeezing direction $\mathbf{\sigma}$, $\Delta\theta=(\Delta{\mathbf{S}}_{\mathbf{\sigma}})^2/|\average{\hat{\mathbf{S}}}|$.
This needs to be compared to the angular noise in the angular momentum coherent states, $1/\sqrt{N}=1/\sqrt{2S}$.
Hence, the squeezing parameter measuring the sensitivity of the squeezed states considered before ($\average{{\hat S}_{2}}=\average{{\hat S}_{3}}=0$) under rotations around the $2$-axis reads
\begin{equation}
\label{eq:entanglement}
\xi_{R}^2=\frac{N(\Delta S_{3})^2}{\average{S_{1}}^2+\average{S_{2}}^2}.
\end{equation}
The area allowed by $\xi_{R}\le1$ is indicated in \Fig{PhaseDiagIntro} by dark shading.

In summary, the fluctuations of the spin in one direction have to be reduced below shot noise ($(\Delta S_{3})^2<S/2$), and the spin polarization in the orthogonal plane, $\average{S_{1}}^2+\average{S_{2}}^2$, has to be large enough to maintain the sensitivity of the interferometer. 
The precision of such a quantum-enhanced measurement is $\xi_{R}/\sqrt{N}$ \cite{Sorensen2001a}, whereas the standard quantum limit set by shot noise is $1/\sqrt{N}$.

\section{Dynamics of the production of squeezed states}
\label{sec:Dynamics}

Tuning the barrier height can be employed to produce many-body states with squeezing in the relative number difference of particles in the two wells. 
Such squeezing was observed \cite{Esteve2008a} in the distribution of atoms counted after high-resolution imaging of the atom cloud in subsequent runs of the experiment.
In each such run, a condensate was prepared in the potential \eq{potential} with a low barrier height 
allowing the atoms to be delocalised across the almost flat potential floor. 
Then, the potential barrier was slowly raised, allowing for an almost adiabatic adjustment of the system's state to the modified external conditions.
As anticipated, the finally strong barrier was observed to suppress the fluctuations $(\Delta n)^2$ in the particle number difference below the estimated classical variance $(\Delta n)^2/4\sim N=\overline{n_1+n_2}$. 
This suppression was interpreted \cite{Esteve2008a} as due to the squeezing that manifests itself theoretically in the approximate low-energy many-body eigenstates. 
As we describe in more detail in the following, the near-adiabatic ramp-up of the barrier allows the atoms essentially to remain in the initially populated states.
The change of parameters in the Hamiltonian during the ramp deforms the populated states, causing adiabatic cooling of the system and squeezing, i.e., reduced relative number fluctuations which reflect the localisation of the particles in either of the wells.
In a second stage when the populated levels successively become pairwise quasi-degenerate, isothermal evolution is observed during which the squeezing remains stationary.

Besides a description of the dynamics observed in the experiment \cite{Esteve2008a}, we characterize the common properties as well as the differences between quantum and classical statistical many-body evolution of the system.
To this end we study the dynamical evolution both by direct integration of the von Neumann equation as well as by simulation in terms of a classical field equation of motion derived from the classical Hamiltonian function corresponding to  the operator \eq{BHeqn}.

\subsection{Quantum evolution}
\label{sec:quantumevolution}
We consider a system whose dynamics is described by the Hamiltonian  \eq{BHeqn}.
At the initial time $t=t_{0}$ the gas is assumed to be evenly distributed among the wells of the potential, such that $\average{\hat S_{3}}=0$ at all times $t>t_{0}$. 
Its initial state is described by a canonical density matrix with a given temperature $T_{0}$, with the spectrum determined by diagonalizing  \eq{BHeqn} for a given set of initial parameters $J(t_{0})$ and $U(t_{0})$ in the Rabi regime as discussed in \Sect{spectrum},
\begin{equation}
\label{eq:densitymatrixdef}
  \hat\rho(t_{0})
  =\frac{1}{Z} \sum_{i=0}^N e^{-E_i(t_{0})/k_B T_{0}} \ket{E_i(t_{0})}\bra{E_i(t_{0})},
\end{equation}
with $Z=\sum_i \exp(-E_i(t_{0})/k_B T_{0})$ and Boltzmann's constant $k_B$. 
The time evolution of the Bose-Hubbard parameters $J(t)$ and $U(t)$ is determined by solving the stationary Gross-Pitaevskii equation in the double-well potential for a set of times during a linear ramp-up of the barrier $V_{0}(t)=V_0(t_0)+v_{0}(t-t_{0})$ with $v_{0}/h=(2\pi)2570\,$Hz.
\Fig{UJfig} shows the evolution of $U$ and $J$ during the ramp up of the barrier, starting at $t_{0}=0$, with the time given in units of the duration of the ramp $t_{max}$.

Decreasing in this way the tunneling parameter $J$ and at the same time preserving or increasing the local interactions $U$ drives the system from the Rabi through the Josephson into the Fock regime, see \Sect{spectrum}.
We compute the corresponding time evolution of the system by solving the von Neumann equation for the density operator and study the evolution of the squeezing parameters described in \Sect{squeezing}, in particular of the number variance $(\Delta n)^2=(\Delta S_{3})^2$ and the coherence parameter or relative phase $\alpha=\average{\hat S_{1}}$.
For the purpose of comparing with experimental data we fit the initial temperature $T_{0}$ such that the initial number variance and coherence fit the experimentally determined values. 

\begin{figure}[tb]
\begin{center}
\includegraphics[width=0.45 \textwidth ]{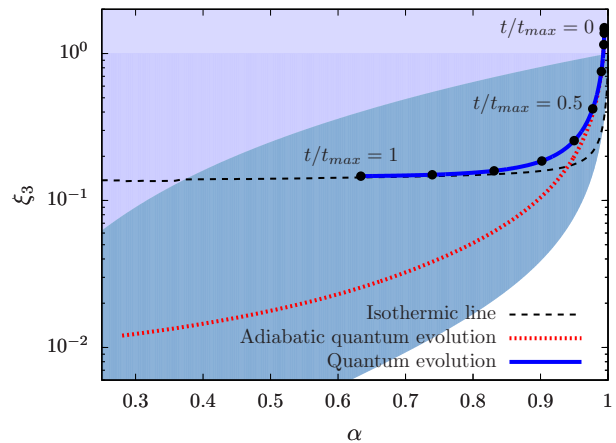}
\caption{ \label{fig:EvolDiag}
(Color online)
Evolution of the coherence $\alpha$ and the number variance $(\Delta n)^2$ for a gas of $N=100$ atoms ((blue) solid line) under a slow ramp-up of $U(t)/J(t)$ over a time $t_{max}=0.16\,$s as given in \Fig{UJfig}.
The system starts in a state with $(\Delta n(t_{0}))^2=37.5$ and $\alpha(t_{0})=0.994$ which is obtained for an initial temperature $T_{0}=20\,$nK.
A range of time points is indicated by black dots. 
They are spaced by $0.1~t_{max}$ and the first completely distinct point is for $t=0.3~t_{max}$.
The evolution under an adiabatic change of the parameters $U$ and $J$, i.e., for $v_{0}\to 0$ is shown by the (red) dotted line.
An isotherm for $T=0.17\,$nK is shown as a dashed line.
See \Fig{PhaseDiagIntro} and \Sect{squeezing} for the definition of the differently shaded areas allowed by the constraint Heisenberg, squeezing, and metrology gain limits.
}
\end{center}
\end{figure}

\Fig{EvolDiag} shows the evolution of a system of $N=100$ atoms in the $\alpha$-$(\Delta n)^2$ plane as a (blue) solid line, under a $t_{max}=0.16\,$s ramp-up of $J(t)/U(t)$ given in \Fig{UJfig}. 
The system starts in a state with $(\Delta n(t_{0}))^2=37.5$ and $\alpha(t_{0})=0.994$ which is obtained for an initial temperature $T_{0}=20\,$nK.
The (red) short-dashed line shows the corresponding evolution of ($\alpha(t)$,$(\Delta n(t))^2$) under an adiabatic change of the parameters $U$ and $J$, i.e., for $v_{0}\to0$.
The areas allowed by the constraint Heisenberg, squeezing, and metrology gain limits, respectively, are shaded differently as discussed in \Sect{squeezing} and \Fig{PhaseDiagIntro}.  

Comparing the nonadiabatic with the adiabatic evolution, as well as with the isothermal line drawn in \Fig{EvolDiag} as a dashed line we find that the system crosses over from an adiabatic to an isothermal evolution.
The transition occurs as soon as the initially occupied levels cross over from the linear to the quadratic part of the spectrum shown for different parameters $U/J$ in \Fig{spectrum}. 

\begin{figure}[tb]
\begin{center}
\includegraphics[width=0.45 \textwidth ]{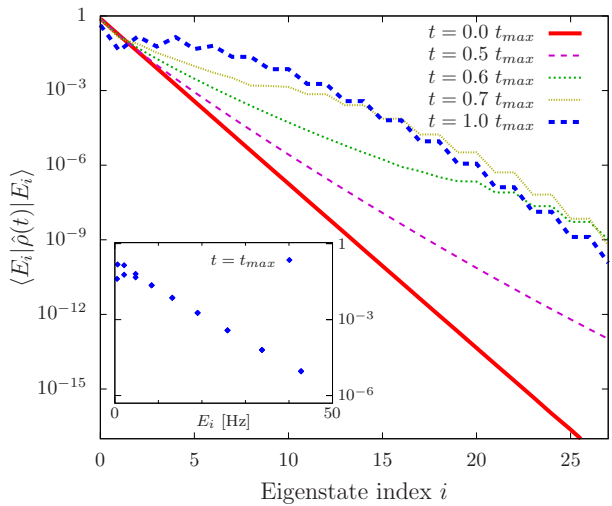}
\caption{ 
\label{fig:DensityMatrix}
Diagonal elements of the density matrix, $\average{E_{i}|\hat\rho(t)|E_{i}}$ in the basis of the energy eigenstates \eq{energyeigenstates} of the Hamiltonian \eq{BHeqn}, for $5$ different times during the ramp-up of $U(t)/J(t)$ shown in \Fig{UJfig}. 
Note the semi-logarithmic scale. The initial temperature is $T_{0}=20\,$nK.
The inset shows the diagonal elements of the density matrix for $t=t_{max}$ as a function of the energy on a logarithmic scale, demonstrating the thermal character of the state.
}
\end{center}
\end{figure}

Let us discuss this crossover in more detail.
We consider the evolution of the density matrix in the energy eigenbasis.
\Fig{DensityMatrix} shows the diagonal elements of the density matrix, $\average{E_{i}|\hat\rho(t)|E_{i}}$ in the basis of the energy eigenstates \eq{energyeigenstates} of the Hamiltonian \eq{BHeqn}, corresponding to the lowest energies $E_{i}$, $i=1,\ldots, 50$.
Note that the density matrix at times $t>0$ is no longer diagonal as correlations have been formed.
Nevertheless, the amplitudes of the off-diagonal elements are strongly reduced compared to the diagonal elements and the latter serve as a measure of the distributions of particles across the energy eigenstates.
\Fig{DensityMatrix} shows these distributions for $5$ different times during the ramp-up of $U(t)/J(t)$ given in \Fig{UJfig}, on a semi-logarithmic scale. 
The linear distribution at $t=0$ corresponds to a canonical density matrix with temperature $T_{0}=20\,$nK, for the linear spectrum in the Rabi regime, see  \Fig{spectrum}. 

Up until around $t=0.5~t_{max}$, the density matrix barely changes its character indicating an adiabatic evolution.
However, as long as the occupied levels are linearly spaced in energy strong adiabatic cooling takes place, as can be inferred from the change in energy scales from the (red) ``Rabi'' to the (blue) ``Josephson'' spectrum in \Fig{spectrum}.
 
Therafter the system quickly leaves its adiabatic behavior to develop a staircase-like shape of the energy distribution.
Around $t \simeq 0.5~t_{max}$ the occupied levels enter the quadratic regime. 
In this regime, the energy spectrum develops two-fold quasi degeneracies, starting in the higher levels, corresponding to symmetric and antisymmetric states, see \Sect{spectrum}.
However, the redistribution between the states in the gradually changing spectrum becomes possible because the levels at the boundary between the linear and quadratic parts of the spectrum approach each other closely.
There is no redistribution within the quasidegenerate pairs in the quadratic part of the spectrum as symmetry forbids transitions between states even and odd in the relative particle number $n$.

For times $t \simeq 0.8~t_{max}$, the energies of all but the largest occupied states belong to the quadratic regime.
Note that at $t=t_{max}$ the occupation of the second lowest energy state, $i=1$, is lower than the occupations of the ground state $i=0$ and second excited state $i=2$.
The reason for this is that the initial state $\hat\rho(t_{0})$ is dominated by the symmetric ground state with $c_{n,0}=c_{-n,0}$ and therefore the final state is predominantly symmetric, with a suppressed contribution from $|E_{1}\rangle$.

We have studied the redistribution of an initial single-energy eigenstate during the ramp and found that the final state is generically far from representing a thermal distribution.
As our results show, however, a thermal mixture of these initial eigenstates redistributes occupation numbers to yield a thermal state again at the endpoint of the ramp.
The inset in  \Fig{DensityMatrix} shows the diagonal elements of the density matrix for the final time $t=t_{max}$ in our simulations, as a function of the energy on a semilogarithmic scale, demonstrating the thermal character of the state in the levels with even index, i.e., the symmetric states, cf. \Fig{Eigenstates}, while the occupation of the odd-$i$ levels remains suppressed.
This suppression stabilizes the system against symmetry breaking to a self-trapped state with a non-zero mean value $n$.
If the odd-$i$ states would be equally strongly occupied, they could combine with the even states to yield self-trapping.

In summary, the system changes considerably during the entire ramp.
During the initial evolution strong adiabatic cooling takes place.
At the crossover to the isothermal evolution redistribution sets in and a ``freeze out'' of the fluctuations, fixing the system's temperature. 
During the following evolution period, the squeezing stays put while the mean coherence keeps decreasing.

\subsection{Semiclassical evolution}
\label{sec:classicalevolution}

We will now turn to the semiclassical statistical description of the production of squeezed states discussed above.
This is achieved by sampling the phase-space probability distribution corresponding to the initial quantum density matrix and evolving each realisation by use of the classical equation of motion.
Correlation functions at a later time are then obtained as moments over the thus propagated probability distribution.
The semiclassical description of the dynamics of the two-mode Bose-Hubbard system has recently been studied in Refs.~\cite{Trimborn2008a}.

The classical dynamic equation \cite{Raghavan1999a} is derived from the classical Hamiltonian function which is obtained from the Hamiltonian in \Eq{HamilAngular} by substituting the operators $\hat S_{1}=\hat S_{+}+\hat S_{-}$ and $\hat S_{3}$ by the classical variables,
\begin{align}
  \hat S_{3}
  &\to n,
  \\
\label{eq:S1phi}
  \hat S_{1} 
  &\to \sqrt{S_{1}^2+S_{2}^2}\cos\phi = \frac{N}{2}\sqrt{1-\frac{4n^2}{N^2}}\cos\phi,
\end{align}
as
\begin{equation}
\label{eq:ClassHamil}
  H =  U n^2 - JN \sqrt{1-\frac{4n^2}{N^2}} \cos\phi. 
\end{equation}
One condition for the classical description to be valid is that the particle number in each well is much larger than one, i.e., $n\ll N$. 
The canonical variables are (half) the number difference $n$ and the relative phase $\phi$ between the two wells.
From the above Hamiltonian the Josephson equations are obtained as
\begin{align}
\label{eq:EqMotion1}
   \frac{dn}{dt}
   &=- N J\sqrt{1-\frac{4n^2}{N^2}}\sin\phi,
   \\
\label{eq:EqMotion2}
   \frac{d\phi}{dt}
   &=2Un-\frac{4 n}{N} J \left(1-\frac{4n^2}{N^2}\right)^{-1/2}\cos \phi.
\end{align}

Expectation values at time $t>t_{0}$ are obtained as moments of the respective probability distribution
\begin{align}
\label{eq:expvalst}
  \average{\cal O}
  &= {\mathcal N}^{-1}\int\, \dd n\, \dd \phi\,P(n,\phi;t)\,\mathcal{O},
  \\
  {\mathcal N}
  &= \int\, \dd n\, \dd \phi\,P(n,\phi;t).
\end{align}
The probability distribution $P$ is determined by the classical path integral
\begin{align}
\begin{split}
\label{eq:classPI}
  P(&n,\phi;t)
  = \int_{t_{0}}^{t}\,{\mathcal D} n\, {\mathcal D} \phi \, P(n,\phi;t_{0})\,\\
  &\times\ \delta[d\phi/dt-2Un-\frac{4 n}{N} J \left(1-\frac{4n^2}{N^2}\right)^{-1/2}\cos \phi],\\
  &\times\ \delta[dn/dt+N J\sqrt{1-\frac{4n^2}{N^2}}\sin\phi]
\end{split}
\end{align}
with the functional measures ${\mathcal D} n=\prod_{\tau=t_{0}}^t \dd n(\tau)$, ${\mathcal D} \phi=\prod_{\tau=t_{0}}^t \dd \phi(\tau)$.
The delta functionals evaluate the variables $n$ and $\phi$ at each point in time according to the solution of the equations of motion, with initial values distributed according to $P(n,\phi;t_{0})$.

For the semiclassical initial state $\hat\rho(t_{0})$,  \Eq{densitymatrixdef}, used in the quantum simulations in the previous section, the probability distribution $P(n,\phi;t_{0})$ at initial time was determined from the Wigner function corresponding to $\hat\rho(t_{0})$.
See Appendix \ref{app:Wigner} for details.

\begin{figure}[tbp]
\begin{center}
\includegraphics[width=0.46 \textwidth ]{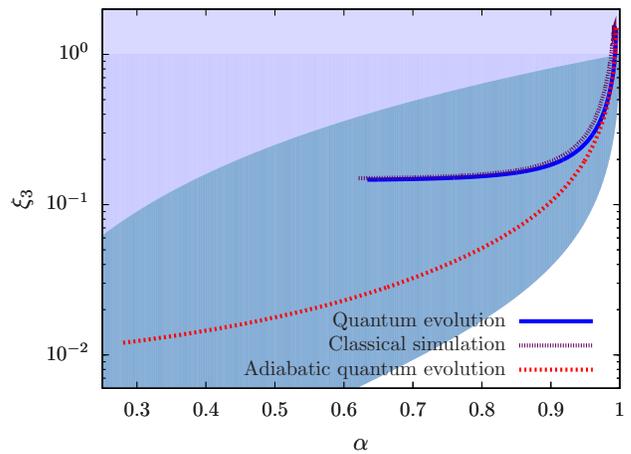}
\caption{ \label{fig:QuantClassComp}
(Color online)
Semiclassical evolution ((purple) dotted line) of the coherence $\alpha$ and the number variance $(\Delta n)^2=N\xi_{3}/4$ for a gas of $N=100$ atoms under a slow ramp-up ($t_{max}=0.16\,$s) of $U(t)/J(t)$ as given in \Fig{UJfig}. 
The (blue) solid line shows the corresponding quantum evolution, as discussed in \Sect{quantumevolution} and \Fig{EvolDiag} for the same evolution time $t_{max}$ and the same time dependent parameters $U$ and $J$.
The initial temperature is $20\,$nK as in \Fig{EvolDiag}.
The definition of the shadings was introduced in \Fig{PhaseDiagIntro}.
}
\end{center}
\end{figure}
\begin{figure}[tbp]
\begin{center}
\includegraphics[width=0.46 \textwidth ]{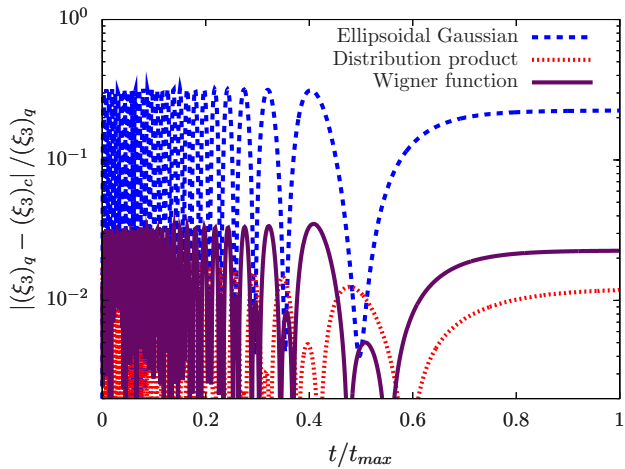}
\caption{ \label{fig:QuantClassCompDeltaxi}
(Color online)
Difference between the $\xi_{3}$ obtained from the semiclassical and the quantum evolutions, normalized to the quantum result, for three different initial phase-space distributions: 
(Purple) solid line: Wigner function for the state \eq{densitymatrixdef}.
(Red) dotted line:  Product of distributions of initial relative number and coherence derived from \Eq{densitymatrixdef}. 
(Blue) dashed line: Ellipsoidal Gaussian state with widths given by these relative number and coherence distributions.
See the main text for a discussion.
}
\end{center}
\end{figure}

\Fig{QuantClassComp} shows the semiclassical evolution of a system of $N=100$ atoms in the previously introduced $\alpha$-$(\Delta n)^2$ plane as a (purple) dotted line, under the ramp-up of $J(t)/U(t)$ given in \Fig{UJfig}. 
The initial classical distribution was calculated from the density matrix \eq{densitymatrixdef} for an initial temperature $T_0=20\,$nK
using the Wigner function \eq{WignerFormula}. 
This distribution is then evolved in time using Eqs.~\eq{classPI}. 
The (blue) solid line shows the corresponding quantum evolution as discussed in \Sect{quantumevolution} and \Fig{EvolDiag} for the same parameters $U(t)$ and $J(t)$, the same initial temperature and the same evolution time $t_{max}$.
Obviously the semiclassical and the quantum evolutions are nearly identical. 
This indicates that dynamics of the production of squeezing is essentially a classical process. 
Nonetheless, the precise shape of the initial probability distribution as derived from the Wigner function corresponding to the state \eq{densitymatrixdef} plays a role as is shown in more detail in \Fig{QuantClassCompDeltaxi} for the same initial temperature
as in \Fig{QuantClassComp}. 
We plot the difference between the $\xi_{3}$ obtained from the semiclassical and the quantum evolutions, normalized to the quantum result, for three different initial phase-space distributions: 
An ellipsoidal Gaussian distribution with main-axes widths given by the number and phase distributions derived from the state \eq{densitymatrixdef} (blue dashed line), 
a product of the distributions of initial relative number and coherence derived from \Eq{densitymatrixdef} (red dotted line), 
and a distribution as given by the full  Wigner function for the state \eq{densitymatrixdef} (purple solid line).
The frequency of the oscillations is approximately given by the plasma frequency in the classical potential and decreases with increasing $U/J$.
We find that the solution derived from the Wigner function shows smaller oscillatory deviations from the exact result than that derived from the Gaussian distribution.
The classical simulations starting from the distribution product give an even smaller deviation.
While the distribution product gives the least deviations during the evolution, the initial-time value of $\xi_{3}$ derived from the Wigner function is closest to the exact result.
Its remaining deviation is due to the discrete sampling of the Wigner function.
We remark that the oscillations take place only during the initial adiabatic decrease of $\xi_{3}$, see \Fig{EvolDiag}, such that they remain mostly invisible in the comparisons shown in \Fig{QuantClassComp}.

The stronger oscillations of the variance of the Gaussian distribution reflect that the initial Wigner function which is almost entirely positive, and therefore classical-like, contains information about non-Gaussian correlations in the initial state.
In particular, at lower temperatures also quantum effects are expected to play a more important role which show up in the negativity of the Wigner function.

\begin{figure}[tbp]
\begin{center}
\includegraphics[width=0.48 \textwidth ]{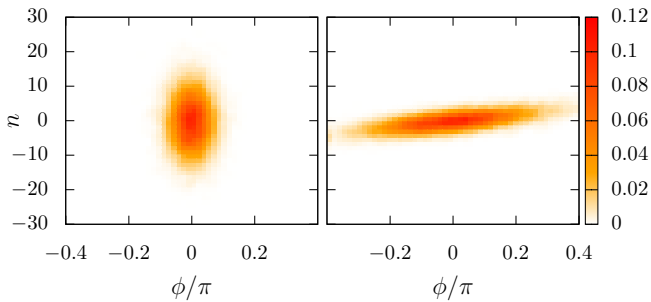}
\caption{ \label{fig:ClassColormap}
(Color online) Evolution of the phase-space probability distribution $P(n,\phi;t)$, \Eq{classPI}, between $t=0$ (left panel) and $t=0.68~t_{max}$ (right panel) as obtained by classical simulations according to \Eq{classPI}. 
White color indicates $P(n,\phi;t)\approx0$ while colors indicate a positive probability. 
The initial distribution is derived from a quantum gas of $N=100$ atoms at a temperature of $T_0=20\,$nK.
The tilt of the final phase-space distributions reflects classical correlations between the generalized position and momentum which reduce the squeezing in $n$ and are due to the non-adiabatic settling to a thermal state, see \Fig{QuantClassComp}. 
}
\end{center}
\end{figure}

In \Fig{ClassColormap}, we illustrate the evolution of the probability distribution $P(n,\phi;t)$, \Eq{classPI}, for $N=100$ particles, for the evolution shown in \Fig{QuantClassComp}.
The left picture shows the initial distribution corresponding to the Wigner function \eq{WignerFormula} at $t=0$ while the right panel shows $P$ after $t=0.68~t_{max}$ of evolution. 
The white areas indicate $P(n,\phi;t)\approx0$ while colors according to the colormap indicate an increasing probability.
%
%
A wide distribution in either direction reflects large fluctuations of the respective observable.
Given an initial distribution this would remain unchanged if the parameter $U$ and $J$ remained constant.
Changing, however, these couplings as in the ramp defined by \Fig{UJfig}, i.e., decreasing the $\phi$-dependent potential term in the Hamiltonian \eq{ClassHamil}, the distribution $P$ varies as the finite distribution over ``momenta'' $n$ leads to an expansion of the distribution in the widened cosine potential, see \Eq{ClassHamil}.
Tuning $J$ to zero allows infinite expansion in the ``position'' direction $\phi$ within the $n$-$\phi$ phase space.
Due to the initial finite distribution in $n$, however, classical correlations between $n$ and $\phi$ develop, tilting the large-time probability distribution with respect to the vertical axis as seen in the right panel of \Fig{ClassColormap} and keeping the expansion finite.
In summary, the evolution of the system in the experiment \cite{Esteve2008a} can be understood to a good approximation as classical squeezing of the phase-space distribution.

\subsection{Quantum statistical squeezing}
\label{sec:QSqueezing}
In the evolution of a weakly interacting quantum gas, quantum fluctuations to leading order enter through zero-point fluctuations in the initial state, i.e., they characterize the scattering into empty and out of nearly empty modes and play little role in the scattering in and out of strongly occupied modes. 
In the language of the path integral this means that the full quantum evolution is, to a good approximation, given by a classical Liouvillean propagation of the initial-time Wigner function which accounts for quantum fluctuations in the initial state, see, e.g., \cite{Berges:2007ym,Polkovnikov2009a}.
Strong interactions have the potential to alter this semi-classical evolution considerably.
However, as is illustrated by our above results, quantum fluctuations also in this case have only little effect if all available modes are strongly occupied during the evolution.

To see distinct effects of quantum fluctuations arising during the evolution requires a larger number of degrees of freedom part of which should stay weakly occupied.
Alternatively, one needs to measure observables with a resolution at the few-particle level.
In the two-well potential the effect of quantum fluctuations may be seen by introducing a tilt such that on the average only a few particles occupy one of the wells.

In the following we stay with equal populations in the two modes but consider the variance of the relative particle number at very low temperatures, where fluctuations on the order of a few atoms become relevant.
At the low temperatures to be considered the initial state is dominated by the ground state of the Hamiltonian.
Driving the system into the Fock regime allows to move to the lower left corner in the graph shown in \Fig{PhaseDiagIntro}. 
For an even total number of atoms the relative number fluctuations between the modes in this regime are strongly reduced while the undefined phase allows interferences on the surface of the Bloch sphere.

\begin{figure}[tb]
\begin{center}
\includegraphics[width=0.45 \textwidth ]{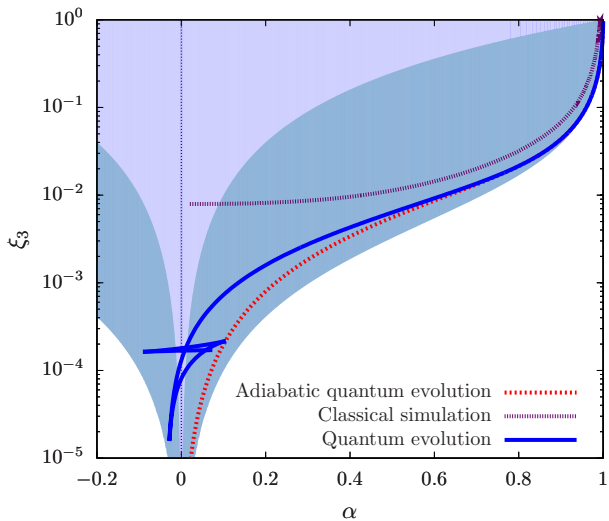}
\caption{ \label{fig:GSDiag}
(Color online)
Evolution of the coherence $\alpha$ and the squeezing parameter $\xi_{3}$ related to the number variance through $(\Delta n)^2=N\xi_{3}/4$ for a gas of $N=100$ atoms under a very slow ramp-up of $U(t)/J(t)$ as given in \Eq{UJdef-gs}. The quantum evolution is represented by the (blue) solid line while the classical statistical evolution is the (purple) dotted line.
The system starts in the ground state of the Bose-Hubbard Hamiltonian for $NU/J(t_{0})\simeq 0.16$.
This corresponds to $(\Delta n(t_{0}))^2=24$ and $\alpha(t_{0})=0.993$. 
The total evolution time is $t_\mathrm{max}=10.18\,$s.
The evolution under an adiabatic change of the parameters $U$ and $J$, i.e., for $1/\tau\to0$ is shown by the (red) short-dashed line.
See \Fig{PhaseDiagIntro} and \Sect{squeezing} for the definition of the areas allowed by the constraint Heisenberg and metrology gain limits, distinguished by different shading.
}
\end{center}
\end{figure}

\Fig{GSDiag} compares the evolution of the exact quantum ((blue) solid line) and semiclassical ((purple) solid line) evolutions of a system starting in the ground state of the Hamiltonian with the ratio of chemical potential over tunneling rate being $NU(t_{0})/J(t_{0})
\simeq 0.16$, for $N=100$ atoms.
This corresponds to $(\Delta n(t_{0}))^2=24$ and $\alpha(t_{0})=0.993$. 
We chose $U=\,$const. and an exponential ramp of 
\begin{equation}
\label{eq:UJdef-gs}
 J(t)/UN=J(t_{0})/UN\exp\{-t/\tau\} 
\end{equation}
with $J(t_{0})/UN=6$ and $\tau=0.55\,$s over the period of $t\lesssim70\,$s.
For a fully adiabatic change of the parameters $U$ and $J$, i.e., for $1/\tau\to0$, the evolution would follow the (red) short-dashed line which corresponds to the dependence of $(\Delta n)^{2}$ on $\alpha$ in the ground state shown as a (red) solid line in \Fig{PhaseDiagEEigenstates}. 

\begin{figure}[tb]
\begin{center}
\includegraphics[width=0.24 \textwidth ]{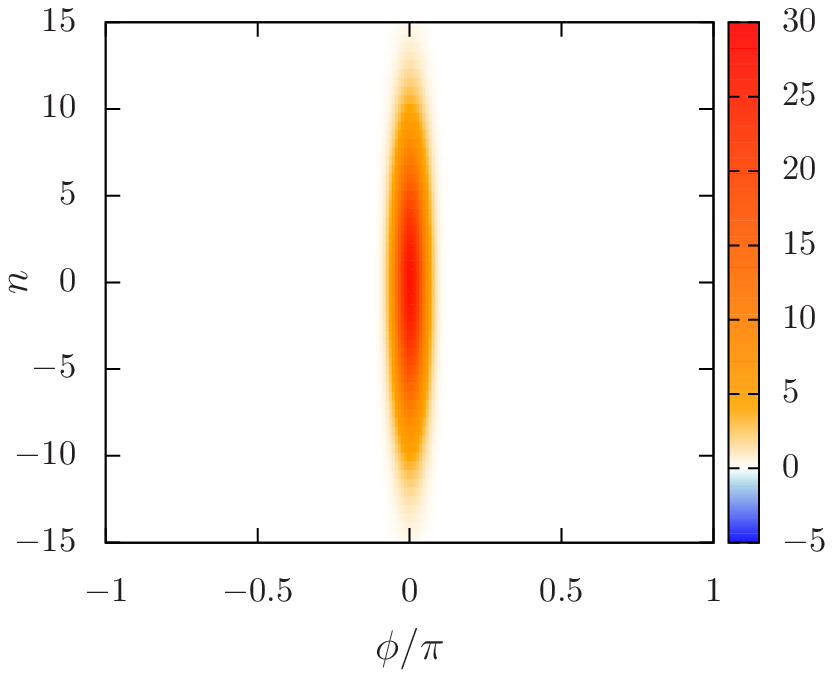}\!
\includegraphics[width=0.233 \textwidth ]{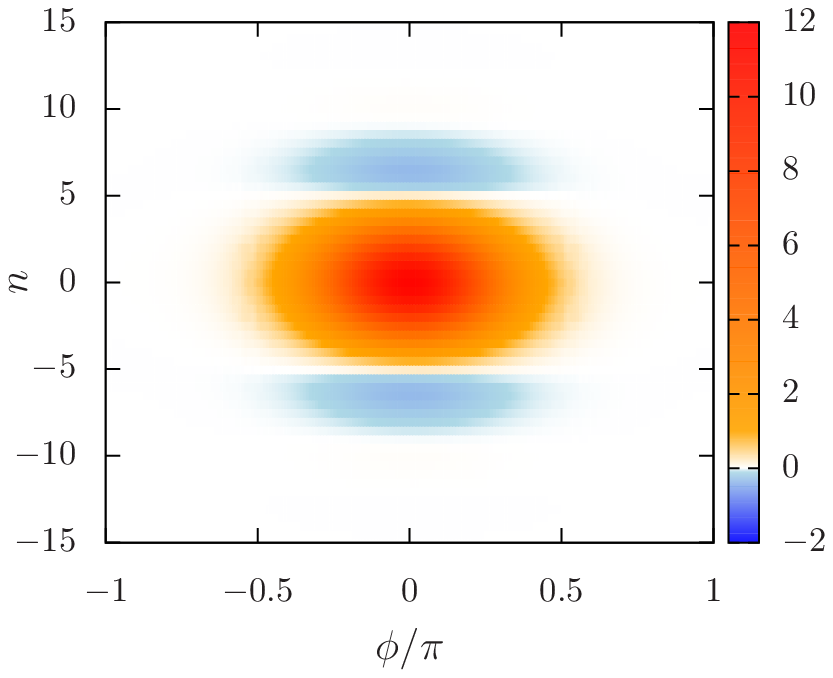}\\
\includegraphics[width=0.24 \textwidth ]{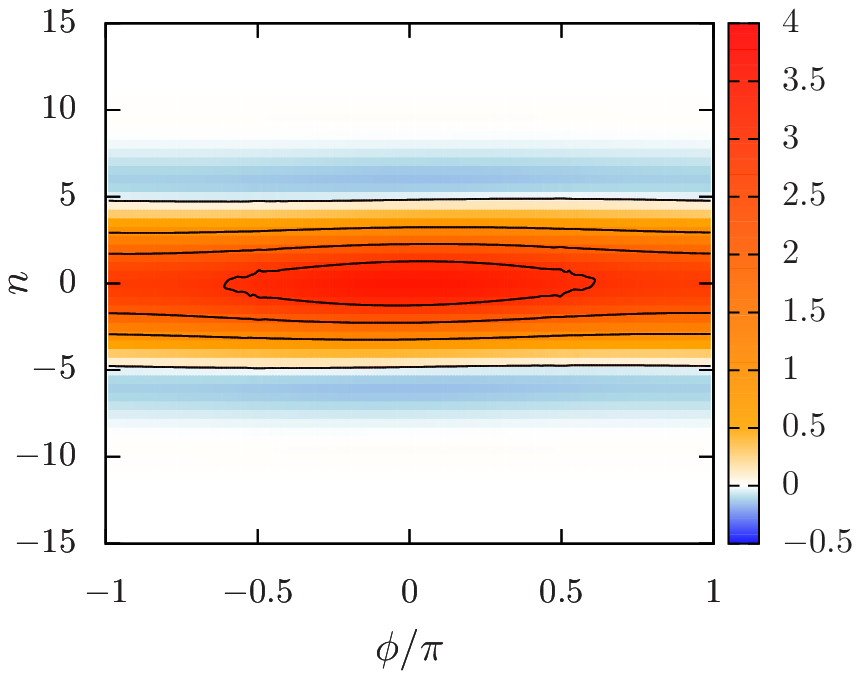}\!
\includegraphics[width=0.24 \textwidth ]{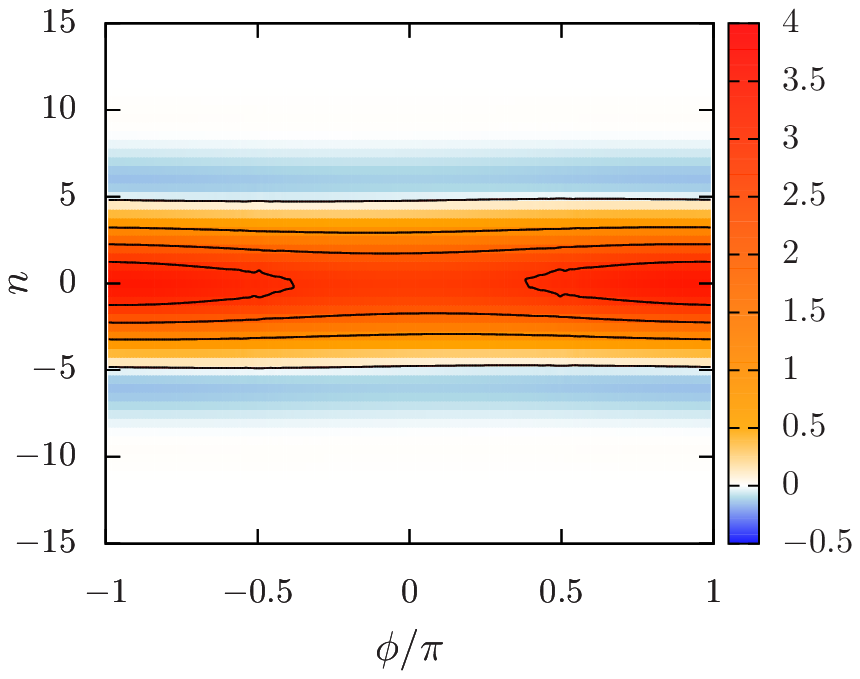}
\caption{ \label{fig:WignerQuantum}
(Color online)
Evolution of the Wigner function for an inverse ramp speed $\tau=0.55\,$s, corresponding to the quantum evolution shown in Figs.~\fig{GSDiag} and \fig{GSFluct}.
Colors encode the value of $W(n,\phi)\times10^{3}$, see \Eq{WignerFormula}, with $n=(|\alpha|^{2}-|\beta|^{2})/2$ and $\phi=2\,\mathrm{arg}(\alpha)$.
$W$ is negative in the blue areas.
The initial distribution shown in the upper left panel is derived from the ground state of the Hamiltonian with $J(t_{0})/NU= 6$ for $N=100$ atoms. 
The upper right panel shows the Wigner function for the number-squeezed state reached at $t/t_{max}=0.5$, with $\xi_{3}(0.5~t_{max})\simeq3\cdot10^{-2}$, $\alpha(0.5~t_{max})\simeq0.86$.
In the lower panels we show the Wigner function at $t/t_{max}=0.75$ (left) and $t/t_{max}=0.9$ (right), i.e. for maximally negative $\xi_{3}\simeq2\cdot10^{-5}$ at $\alpha\simeq{-0.02}$, and maximally negative $\alpha\simeq{-0.1}$ at $\xi_{3}\simeq2\cdot10^{-4}$, respectively (cf.~\Fig{GSDiag}). 
The contour lines help to show the position of the maxima in $|\phi|=0$ and $|\phi|=\pi$, respectively.
}
\end{center}
\end{figure}
\Fig{GSDiag} exhibits significant differences as compared to the semi-classical evolution shown in \Fig{QuantClassComp} which can be expressed both in terms of $(\Delta n)^2$ and $\alpha$.
With respect to atom number fluctuations, the classical curve approaches the value $(\Delta n)^2\simeq0.2$ and cannot follow the quantum one below this limit.
Following the adiabatic evolution for increasing initial temperatures $T$, we find that the quantum regime $(\Delta n)^2\lesssim0.2$ can only be reached for $T\lesssim20\,$nK$\,\simeq10\,UN$.
With respect to the coherence parameter $\alpha$, the classical curve simply approaches zero indicating a distribution on the Bloch sphere symmetric under $S_{1}\leftrightarrow -S_{1}$.
In this limit, the classical phase space distribution wraps around the equator of the Bloch sphere and reflects a completely undetermined relative phase between the wells. 

In contrast, the quantum evolution yields a coherence which oscillates around zero, corresponding to an asymmetric distribution of phases.
This distribution is exhibited by the Wigner function which we show, for the time when $\alpha$ is most negative, in \Fig{WignerQuantum}.
Due to interferences the Wigner function starts to oscillate as soon as the phase distribution fully wraps around the equator.
This leads to oscillations both in $(\Delta n)^2$ and $\alpha$ during the near-adiabatic time evolution shown in \Fig{GSDiag}.
The oscillations in $(\Delta n)^2$ are damped and equilibrate at a value close to the point where $\alpha$ reached zero for the first time.
In the lower panels of \Fig{WignerQuantum} the Wigner function is shown for the evolution times $t/t_{max}=0.75$ (left) and $t/t_{max}=0.9$ (right), i.e. for maximally negative $\xi_{3}\simeq2\cdot10^{-5}$ at $\alpha\simeq{-0.02}$, and maximally negative $\alpha\simeq{-0.1}$ at $\xi_{3}\simeq2\cdot10^{-4}$, respectively (cf.~\Fig{GSDiag}).
This shows that negative $\alpha$ arises from a maximum of the Wigner function at $n=0$, $|\phi|=\pi$. 
Moreover, in this pure quantum regime of the evolution, $W$ becomes negative, for all $\phi$, periodically in $n$, for $|n|>5$.
The first minima at $|n|\simeq6$ are at the center of the grey zones.

\begin{figure}[tbp]
\begin{center}
\includegraphics[width=0.45 \textwidth ]{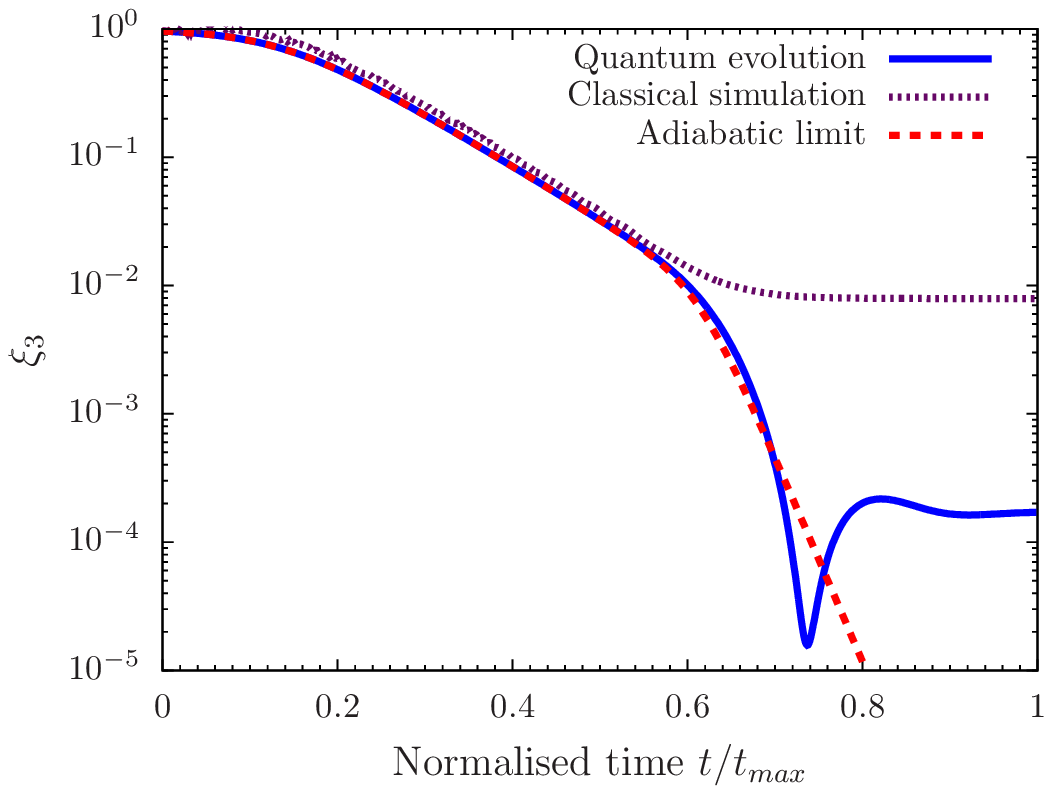}
\caption{ \label{fig:GSFluct}
(Color online)
Evolution of the squeezing parameter $\xi_{3}$ related to the number variance through $(\Delta n)^2=N\xi_{3}/4$ for a gas of $N=100$ atoms under a very slow ramp-up of $U(t)/J(t)$ as given in \Eq{UJdef-gs}. The quantum evolution is represented by the (blue) solid line while the classical statistical evolution is the (purple) short-dashed line.
The initial state and all other parameters are chosen as in \Fig{GSDiag}.
The evolution under an adiabatic change of the parameters $U$ and $J$ is shown by the (red) long-dashed line.
}
\end{center}
\end{figure}

As was shown in Ref.~\cite{Javanainen1999b} the relative number variance $(\Delta n)^{2}$ in the ground state as a function of the ratio $NU/4J$ of chemical potential over tunneling coupling, has the approximate value 
\footnote{Note that in Eqs.~(18), (23), and (25) of Ref.~\cite{Javanainen1999b}, the factor $\sqrt{N/2}$ should rather read $\sqrt{N}/2$.}
\begin{equation}
\label{eq:DeltanGSAdiab}
   \xi_{3}=4(\Delta n)^{2}/N=(1+NU/4J)^{-1/2},
\end{equation}
in the Josephson regime where the variance is distinctly smaller than the classical limit, $\xi_{3}\ll1$, but sufficiently larger than $1$, i.e., where $N^{-1}\ll U/J \ll N$, see \Sect{spectrum}.

\Fig{GSFluct} shows the dependence of $\xi_{3}=4(\Delta n)^{2}/N$ on $t$, on a semi-logarithmic scale, as above for the exponential ramp \eq{UJdef-gs} with $\tau=0.55\,$s.
The (red) dashed line corresponds to adiabatic evolution ($\tau\to\infty$) while the (blue) solid and (purple) short-dashed lines represent the near-adiabatic quantum and semiclassical evolutions for $\tau=0.55\,$s, respectively.
The adiabatic curve shows three regimes in each of which $\log\xi_{3}$ varies linearly with time which correspond to the Rabi, Josephson, and Fock regimes.
The full quantum evolution as shown in \Fig{GSFluct} exhibits an oscillatory behavior of $\xi_{3}$, during which the squeezing parameter undershoots the adiabatic ground-state curve and eventually settles to a finite value smaller than that in the classical limit.
Although $\xi_{3}$ falls below the value it can reach in an adiabatic ramp, this does not contradict the Heisenberg limit as is clearly seen in \Fig{GSDiag}.

\begin{figure}[tbp]
\begin{center}
\includegraphics[width=0.45 \textwidth ]{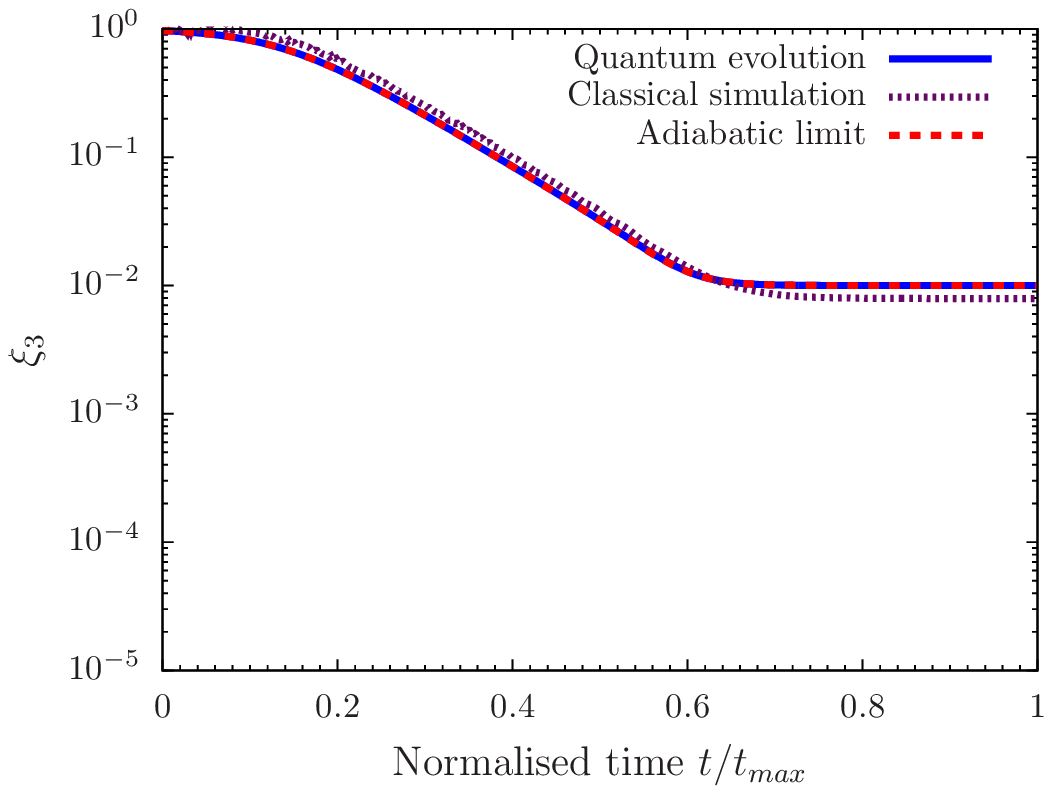}
\caption{ \label{fig:GSOdd}
(Color online)
Evolution of the squeezing parameter $\xi_{3}$ related to the number variance through $(\Delta n)^2=N\xi_{3}/4$ for a gas of $N=101$ atoms under a very slow ramp-up of $U(t)/J(t)$ as given in \Eq{UJdef-gs}. 
The quantum evolution is represented by the (blue) solid line while the classical statistical evolution is the (purple) short-dashed line.
All other parameters are chosen as in \Fig{GSFluct}.
The adiabatic evolution is again shown by the (red) long-dashed line.
}
\end{center}
\end{figure}

As a contrast we show, in \Fig{GSOdd}, that squeezing below the classical limit is not possible for systems with an odd total number $N$ of particles.
In this case, the semiclassical evolution perfectly describes the dynamics of the formation of squeezing correlations.
To see the difference to the case of an even number of particles one trivially has to measure the particle number to better than half a particle.

Taking into account the time evolution \eq{UJdef-gs}, our data confirms the approximate expression \eq{DeltanGSAdiab} in the Josephson regime.
In the Fock regime the dependence of $\xi_{3}$ on $J$ is approximately given by \cite{Javanainen1999b}
\begin{equation}
\label{eq:DeltanGSAdiabFock}
   \xi_{3}=4(\Delta n)^{2}/N=4NJ^2/U^2,
\end{equation}
and the transition between the Josephson and Fock regimes occurs where $J/U\simeq 2^{-4/3}N^{-1}$.

\begin{figure}[tbp]
\begin{center}
\includegraphics[width=0.45 \textwidth ]{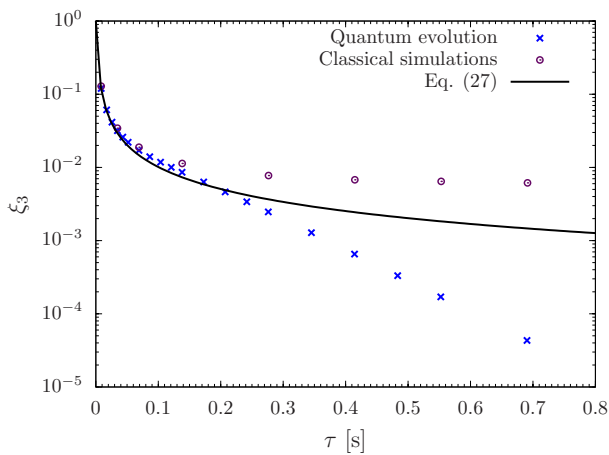}
\caption{ \label{fig:xi3limit}
(Color online)
Final value of $\xi_{3}(t_{\mathrm{max}})$ obtained after an evolution time $t_{\mathrm{max}}=18.4\tau$ vs ramp speed $\tau$.
The solid line corresponds to the analytical result in \Eq{xi3limit} valid in the Rabi and Josephson regimes.
The (blue) crosses and (purple) circles are the corresponding final values obtained from a full quantum evolution and semiclassical evolution of the initial state, respectively.
In the Fock regime which can be reached for ramp rates $1/\tau<2\pi\,$s$^{-1}$ the squeezing becomes stronger than the classical limit $\xi_{3}\simeq6\cdot10^{-3}$.
}
\end{center}
\end{figure}

Considering the non-adiabatic quantum evolution for $J(t)/UN=(J(t_{0})/UN)\exp\{-t/\tau\}$ we find that $\xi_{3}(t)$ follows the ground-state dependence as long as the ramp rate is smaller than the Josephson frequency \eq{omegaplasma}, $1/\tau\ll\omega_{p}$.
As soon as the decreasing frequency $\omega_{p}$ falls below $1/\tau$ the squeezing parameter $\xi_{3}$ is frozen out at the approximate value \cite{Javanainen1999b}
\begin{equation}
\label{eq:xi3limit}
   \xi_{3}=\sqrt{1+(\pi NU\tau/2)^{2}}-\pi NU\tau/2.
\end{equation}
\Fig{xi3limit} shows this limit as a function of the inverse ramp rate $\tau$.
The solid line gives the analytical formula \eq{xi3limit} while the (blue) crosses and (purple) circles correspond to the asymptotic values for $\xi_{3}$ determined from a set of our quantum and semiclassical evolutions, respectively.
Clearly, the semiclassical evolution cannot enter the Fock regime.

\section{Conclusion}
\label{sec:Concl}
We have studied in detail the production of squeezed states in an ultracold Bose gas in a double-well trap.
The trapping parameters are chosen as in the experiment \cite{Esteve2008a} such that at the temperatures considered, the system can be described by a two-site Bose-Hubbard Hamiltonian.
Following the experimental procedure, the gas is initially confined by a double-well trap with very weak interwell barrier such that free tunneling is possible between the sites.
We studied the time evolution of the system under a slow but non-adiabatic ramp-up of the barrier, in particular with respect to the change in the variance of the particle number difference between the sites and the coherence which is related to the expectation value of the relative phase.
In this way a many-body state with squeezing in the particle number difference, i.e., reduced variance of this observable at the expense of the variance of the relative phase, is prepared.
Our results confirm that the squeezing attainable with a finite barrier ramp-up speed is limited to a value depending on the initial temperature and the ramp speed.
This dependence is determined by the spectrum of the model Hamiltonian in which the low-energy states become quasi-degenerate below a certain ratio of the tunneling rate over the on-site energy.
Once the tunneling is sufficiently suppressed such that the two lowest states are separated by a frequency on the order of the inverse ramp rate, the squeezing saturates.

We have formulated the model and dynamic equations in terms of Bloch angular momentum operators and their correlation functions to obtain a pictorial description of the underlying dynamics and exhibit the connection to spin squeezing.
Beyond a qualitative understanding of the experimental data of Ref.~\cite{Esteve2008a} our focus was set on the distinction between quantum and classical statistical fluctuations.
For this, classical statistical simulations were conducted and compared to the full quantum evolution and description in terms of the Wigner function.
Our results show that within the parameter regime realized in the experiment, the production of squeezing is an entirely classical process.
Squeezing below the classical limit is possible, however, due to the low degree of degeneracy in the system, only at significantly lower temperatures than in experiment.
The detection of such squeezing and quantum correlations requires the measurement of particle number at the single-particle level.
Crucial differences arise for systems with an even total particle number as compared to such with an odd number.
The results are readily applicable to other realisations which can be described by the model employed.
We have shown that in the regime where quantum fluctuations become relevant, maximum squeezing is in general not achievable starting from mixtures of energy eigenstates and invoking adiabatic parameter changes.
We emphasize that our results may be particularly interesting for mesoscopic dynamics experiments.

\acknowledgments \noindent 
The authors would like to thank S. Giovanazzi, C. Gross, M. Holland, C. Kollath, M. Kronenwett, and D. Meiser for inspiring and useful discussions. 
C.\,B.\ and T.\,G.\ would like to thank M. Holland, JILA, and the University of Colorado for their hospitality, and acknowledge support by the Deutsche Forschungsgemeinschaft  as well as by the Alliance Program of the Helmholtz Association (HA216/EMMI).
C.\,B.\ thanks the Heidelberg Graduate School for Fundamental Physics (HGSFP) and the Landesgraduiertenf\"orderung Baden-W\"urttemberg for support.

\begin{appendix}

\section{Bose-Einstein condensate in a double-well potential}
\label{app:StatSys}
\begin{figure}[tbp]
\begin{center}
\includegraphics[width=0.4 \textwidth ]{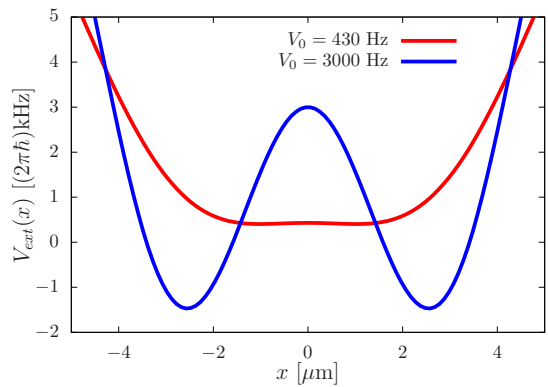}
\caption{ \label{fig:doublewell} 
Trapping potential $V_\mathrm{ext}$ realised in the experiment \cite{Esteve2008a} for two
different barrier heights before and after an adiabatically conducted ramp-up.
}
\end{center}
\end{figure}

We consider a Bose condensate trapped in a double-well potential with two energetically
degenerate minima separated by a barrier of variable height, allowing for an
adjustable tunneling rate between the wells.
In the experiment \cite{Esteve2008a} such a potential was formed
optically by counterpropagating laser waves creating the superposition of
standing waves with two different frequencies. Near the trapping minima the
potential can be approximately described as
\begin{equation}
\label{eq:potential}
V_\mathrm{ext}(x)=V_0 \cos(kx)+\frac{1}{2}m\omega^2 x^2
\end{equation}
\Fig{doublewell} shows the potential form for two different
barrier heights realized in the experiment \cite{Esteve2008a}. 

In the following we introduce the many-body Hamiltonian
and review its properties in the parameter regime chosen in the experiment
\cite{Esteve2008a}. 
We choose the description in terms of a two-mode model which is applicable at the low temperatures encountered in the experiment. 
For this model, the Hamiltonian can be diagonalised exactly and the number as well as the conjugate phase distribution be studied accordingly.

\subsection{Hamiltonian}
\label{subsec:sba}
As in \cite{Esteve2008a}, we consider a system of identical bosons, of mass $m$, trapped in the one-dimensional potential $V_\mathrm{ext}(x)$,  \Eq{potential}. It is described by the Hamiltonian
\footnote{Where not otherwise mentioned we choose natural units where $\hbar=1$.}
\begin{equation}
\begin{split}
\label{eq:Hamil2}
 & \hat H=\int \dd x~\left\lbrace \hat\Psi^\dagger(x) \left(  -\frac{\partial_x^2}{2 m}+V_\mathrm{ext}(x) \right) \hat\Psi(x)\right.  \\
   & \left. \phantom{\left(  \frac{\partial_x^2}{2} \right)} + \frac{g}{2} \hat\Psi^\dagger(x)\hat\Psi^\dagger(x) \hat\Psi(x)\hat\Psi(x)
  \right\rbrace .
\end{split}
\end{equation}
Here the binary interactions between the particles are modeled by an effective contact potential $V_\mathrm{int}(x)=g\delta(x)$, with the coupling constant being proportional to the s-wave scattering length $a$, $g=4\pi a / m$. 

Within the single-particle sector, the ground and first excited states of the Hamiltonian \eq{Hamil2}, in the absence of a barrier ($V_0=0$), are the corresponding harmonic oscillator eigenstates. 
Raising the barrier adiabatically these states are transformed into the lowest two energy eigenstates $\psi_0(x)$, $\psi_1(x)$ in the double well, both with amplitudes peaked within the two wells. 
As for the harmonic oscillator states, the ground (excited) state is (anti-)symmetric under the reflection $r\rightarrow-r$ (see \Fig{doublewell}).

In the experiment \cite{Esteve2008a}, the frequency of the harmonic trap was chosen $\omega\simeq (2\pi) 5\,$Hz while the amplitude of the barrier was varied in the range between $V_{0}\simeq (2\pi) 430\,$Hz and $V_{0}\simeq (2\pi) 3\,$kHz.
\Fig{doublewell} shows the potential in these limits.
%
\begin{figure}[tb]
\begin{center}
\includegraphics[width=0.45 \textwidth ]{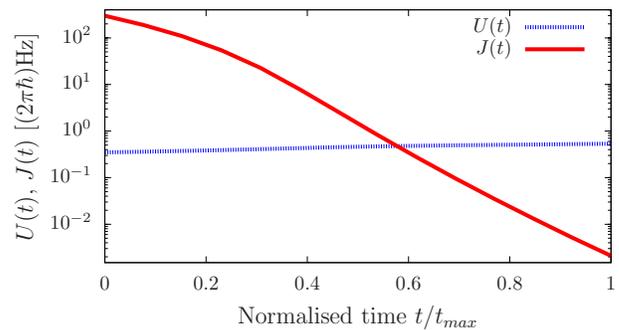}
\caption{ \label{fig:UJfig}
Time evolution of the parameters $U(t)$ and $J(t)$ of the Bose-Hubbard Hamiltonian as realized in the experiment \cite{Esteve2008a}. 
This evolution drives the system from the Rabi to the Fock regime.
}
\end{center}
\end{figure}

For the maximum temperatures reached in \cite{Esteve2008a}, we can restrict ourselves to a description in terms of a two-site Bose-Hubbard Hamiltonian
\begin{equation}
\label{eq:BHeqn-app}
  \hat H 
  = - J (\hat a_1^\dagger \hat a_{2}+\hat a_{2}^\dagger \hat a_1) 
     + \frac{U}{2} \sum_{i=1}^2 \hat a_i^\dagger \hat a_i^\dagger \hat a_i \hat a_i.
\end{equation}
where the operators $\hat a_i$ and $\hat a_i^\dagger$ obey the standard bosonic commutation relations \eq{commutation} and where $J$ is the tunneling and $U$ is the onsite interaction parameter.
Solving the three-dimensional Gross-Pitaevskii equation for the trapping potential realized in \cite{Esteve2008a} one finds the parameters $U$ and $J$ as functions of the barrier height between the wells.
They are shown, for the potential during the near-adiabatic ramp, in \Fig{UJfig}.
For potentials between the limiting cases depicted in \Fig{doublewell} the minimum width of the lowest band of states is obtained in the limit of large $U/J$.
As will become clear in the following subsection, the width in this limit is approximately $UN^2/4\simeq(2\pi)1.2\,$kHz, for a total number of particles $N=100$.
To estimate the validity of the single-band model this needs to be compared to the typical temperatures in the experiment which are on the order of $10^2\,$nK, i.e., $10^{-7}\,$Hz.

\subsection{Energy spectrum}
\label{sec:spectrum}
In the following we limit ourselves to the canonical ensemble, with a fixed total number of
particles $N$.
A convenient basis to express the energy eigenstates of \eq{BHeqn-app} in are
the Fock number eigenstates of $\hat n=(\hat a_{1}^\dagger\hat a_{1}-\hat a_{2}^\dagger\hat a_{2})/2$,
\begin{equation}
\label{eq:fockeigen}
 \hat n \ket{N/2+n,N/2-n}=n \ket{N/2+n,N/2-n},
\end{equation}
with $-N/2\leq n \leq N/2$.
The dimension of the Hilbert space is $N+1$. 
In practice, this allows us to study the energy spectrum by diagonalizing numerically the Hamiltonian. 
For the following illustrations we choose a particle number of $N=100$.

Properties of the Hamiltonian \eq{HamilAngular} and the fragmentation of a Bose-condensate in a double-well potential have been discussed in detail before \cite{Cirac1998a,Steel1998a,Spekkens1999a,Mahmud2003a, Streltsov2004a}, and we review here only the aspects relevant to our discussion. 
For a general review see Ref.~\cite{Leggett2001a}.
See also Refs.~\cite{Dunningham1999a,Dunningham2001a} 
for proposals for the manipulation of number and phase correlations in condensates trapped in a double well. 
As far as the form of the spectrum is concerned, three different regimes can be distinguished by means of the ratio $U/J$ \cite{Paraoanu2001a}. 
In the \emph{Rabi regime}, $U/J \ll N^{-1}$, the system consists of $N$ nearly independent particles. 
This corresponds to the non-interacting limit. 
In the \emph{Josephson regime}, $N^{-1} \ll U/J \ll N$, atom number fluctuations are small and coherence is high. 
This is called the classical regime as eigenstates are described by predominantly positive Wigner functions with widths near the standard quantum limit. 
The \emph{Fock regime}, $N \ll U/J$, is dominated by the interaction energy $U$, thus the atom number in each well is well defined.
Reduced number fluctuations and other non-classical effects appear in the Fock regime.

\begin{figure}[tb]
\begin{center}
\includegraphics[width=0.45 \textwidth ]{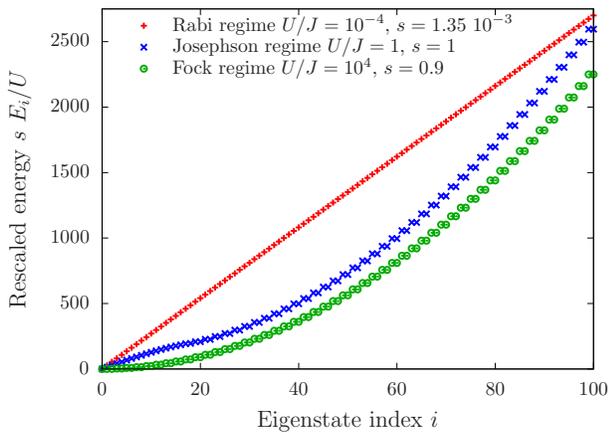}
\caption{ 
\label{fig:spectrum}
Spectrum of the Bose-Hubbard Hamiltonian in different regimes. 
The energies are given in units of $U/s$, with $s$ a scale factor given in the legend. 
All energies are shifted such that $E_0=0$.
}
\end{center}
\end{figure}

In the Rabi regime, the spectrum consists of a series of equally spaced states with a level spacing of $\omega_p$ where
the plasma frequency $\omega_p$ is given by
\begin{equation}
\label{eq:omegaplasma}
   \omega_p
   =\sqrt{2J\left(NU+2J\right)}
\end{equation}
Increasing $U/J$ beyond $1/N$ introduces an approximately quadratic part to the spectrum for energies $E \gtrsim 2NJ$, for which
\begin{equation}
  E_i
  \simeq NJ+\frac{U}{4}i^2.
\end{equation}
In the Fock regime only quasi-degenerate pairs of states remain, forming an approximately quadratic spectrum. 
It was shown in Ref.~\cite{Salgueiro2007a} that the splitting between these quasi-degenerate states vanishes with $1/N!$.

\begin{figure}[tb]
\begin{center}
\includegraphics[width=0.45 \textwidth ]{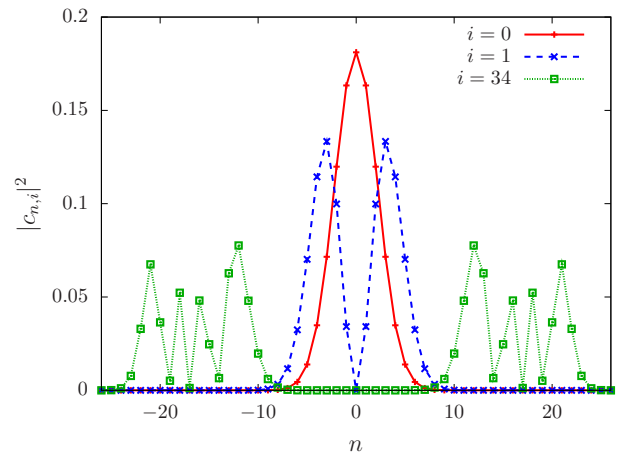}
\caption{ \label{fig:Eigenstates}
Distribution of the relative particle number $n=(n_{1}-n_{2})/2$ in three different eigenstates of the
Hamiltonian in the Josephson regime, $U/J=1$, for $N=100$ particles.
For the states with odd $i$, the coefficients are odd under $n\to-n$, and vice-versa for the even-$i$ states. 
}
\end{center}
\end{figure}

\Fig{Eigenstates} shows the occupation number distribution $|c_n|^2$ of three different energy eigenstates $\ket{E_i}$ in the Fock basis $|n\rangle=|N/2+n,N/2-n\rangle$ defined in \Eq{fockeigen},
\begin{equation}
\label{eq:energyeigenstates}
  \ket{E_i}=\sum_{n=-N/2}^{N/2}c_{n,i}\ket{n}.
\end{equation}
We find that the coefficients of the ground state are well approximated by a Gaussian centered around $n=0$, reflecting its semiclassical nature. 

The first exited state has a wider distribution around the same mean value.
Since it must be an antisymmetric state, one has $c_{1,0}=0$. 
The higher excited states show an increasingly wider distribution around $i=0$ with an increasing number of ``nodes'' in $c_{n,i}$. 

In the Fock regime, $U/J>N$, all states are quasi twofold degenerate. 
In this regime, there is no longer any contribution near $n=0$. 
A gap opens up in the distribution $|c_{n,i}|^2$ which increases towards larger $i$. 
In the limit $U/J\rightarrow\infty$ the highest excited state,
\begin{equation}
  \ket{E_{N+1}}
  =\frac{1}{\sqrt{2}}\left(\ket{N,0}+\ket{0,N}\right).
\end{equation}
is cat like.
Note, however, that excitation of the double-well system up to $\ket{E_{N+1}}$, e.g., at higher temperatures, easily exceeds the validity of the single-band approximation.

\subsection{Angular-momentum representation}
\label{sec:squeezing}
In many-body quantum physics, number and phase are a particular type of conjugate variables which fulfill a Heisenberg-type uncertainty relation. 
We discuss, in this subsection, the representation of the Hamiltonian in terms of number and phase operators and review different measures of squeezing.
The uncertainty relation allows us to derive fundamental limits for the respective squeezing parameters.

The relation between number and phase can be most conveniently visualised on the Bloch sphere, i.e. in terms of the Schwinger representation  \cite{Schwinger1965a} of the SU(2) symmetry group,
\begin{equation}
\label{eq:Jdefine}
\begin{split}
  \hat S_k 
  = \frac{1}{2} \sum_{i,j=1}^{2}\hat a_i^\dagger\sigma^{k}_{ij}\hat a_{j},
  \quad k=1,2,3,
\end{split}
\end{equation}
where $\sigma^{k}$ is the Pauli $k$-matrix.
As a consequence of the commutation relations \eq{commutation}, the operators $\hat S_i$ form the fundamental representation of the angular momentum algebra, with
\begin{equation}
\label{eq:angular}
  [\hat S_k, \hat S_l]
  =i\varepsilon_{klm}\hat S_m,
\end{equation}
$\varepsilon_{klm}$ being the total antisymmetric tensor of rank 3.
These commutation relations give rise to a set of uncertainties relations that determine lower bounds for the
fluctuations of the observables,
\begin{align}
  (\Delta S_{k})^2(\Delta S_{l})^2 \ge \frac{1}{4} |\epsilon_{klm}\langle \hat S_{m}\rangle|^{2}.
  \nonumber
  \tag{\ref{eq:UncertaintyS}'}
\end{align}
with $(\Delta S_{k})^2=\average{(\hat S_k-\average{\hat S_k})^2}$.
In this representation the two-mode states with fixed total particle number $N$ can be written as angular
momentum states $\ket{S,S_3}$ with $S=N/2$, $S_3=n$.
The Hamiltonian \eq{BHeqn-app} in this representation is given in \Eq{HamilAngular}.

\section{Wigner function of two-mode states}
\label{app:Wigner}
The Wigner function of a two-mode system can be calculated from its definition
\begin{equation}
\label{eq:wignerdef}
W(\alpha,\beta)=\frac{1}{\pi^4}\int\dd^2\lambda\,\dd^2\mu\, C_S(\lambda,\mu)
                e^{\alpha \lambda^\star-\alpha^\star\lambda}
                e^{\beta\mu^\star-\beta^\star\mu},
\end{equation}
where $C_S$ is the symmetrically ordered characteristic function, which one obtains as follows \cite{Barnett1997a}.
See also Ref.~\cite{Dowling1994a} for a detailed discussion of two-mode Wigner functions on the Bloch sphere.
First, the $Q$-function
\begin{equation}
Q(\alpha,\beta)=\frac{1}{\pi^2}\bra{\alpha,\beta}\hat\rho\ket{\alpha,\beta}
\end{equation}
is determined as the expectation value of the density matrix $\hat\rho$ with respect to the two-mode coherent state
\begin{equation}
\ket{\alpha,\beta}=e^{-(|\alpha|^2+|\beta|^2)/2}\sum_{i,j=0}^\infty 
\frac{\alpha^i\beta^j}{\sqrt{i!j!}}\ket{i}_l\ket{j}_r.
\end{equation}
The antinormally ordered characteristic function is then obtained by 
Fourier transforming the $Q$-function,
\begin{equation}
  C_A(\lambda,\mu)
  =\int\dd^2\alpha\,\dd^2\beta\,
  Q(\alpha,\beta)\,
  e^{\lambda\alpha^\star-\lambda^\star\alpha}
  e^{\mu\beta^\star-\mu^\star\beta}.
\end{equation}
The symmetrically ordered characteristic function is finally calculated using the Baker-Campbell-Hausdorff relation, leading to
\begin{equation}
  C_S(\lambda,\mu)
  =C_A(\lambda,\mu)\, e^{(|\lambda|^2+|\mu|^2)/2},
\end{equation}
and from this the Wigner function by use of \Eq{wignerdef}.

Applying this procedure and using the representation of the density matrix in the Fock basis \eq{fockeigen}, $|n\rangle=|N/2+n,N/2-n\rangle$,
\begin{equation}
  \rho_{nm}
  = \langle n|\hat\rho|m\rangle,
\end{equation}
we arrive at
\begin{equation}
\begin{split}
\label{eq:WignerFormula}
  W(\alpha,\beta)
  =&\frac{4^{1-N}}{\pi^2}\sum_{n,m=0}^N
  \frac{\rho_{nm}}
       {\sqrt{n!m!(N-n)!(N-m)!}}\\
  & \times\ \Omega_{nm}(\alpha)\Omega_{N-n~N-m}(\beta) e^{-2(|\alpha|^2+|\beta^2|)},
\end{split}
\end{equation}
with
\begin{equation}
\begin{split}
  \Omega_{nm}(\alpha)
  =&\sum_{k=0}^n  \binom{n}{k} \sum_{l=0}^m \binom{m}{l}(-1)^k i^{k+l} \\ 
  & \times\ H_{k+l}(2\,\text{Im}\alpha) H_{n+m-(k+l)}(2\,\text{Re}\alpha).
  \label{eq:Omeganm}
\end{split}
\end{equation}
where $H_n(x)$ is the $n$th Hermite polynomial.
Eqs.~\eq{WignerFormula} and \eq{Omeganm} show that $W$ is real if $\rho$ is hermitian.

As we only consider closed systems, the dependence of $W$ on the absolute phase of the two modes is irrelevant. 
We therefore need to evaluate $W$ only for different relative phases $\phi$, choosing, e.g., 
\begin{equation}
  \alpha
  =|\alpha|e^{i\phi/2}\text{,~~~~~~}\beta=|\beta|e^{-i\phi/2}.
\end{equation}
Taking furthermore into account that the total number of particles $N=|\alpha|^2+|\beta|^2$ is fixed reduces the number of free arguments of $W$ to two, the quantities $n=(|\alpha|^{2}-|\beta|^{2})/2$ and $\phi=2\mathrm{arg}(\alpha)$.

\end{appendix}

\bibliography{bibtex/additions,bibtex/mybib}

\end{document}